\documentclass[aps,twocolumn]{revtex4}

\usepackage{amsmath}
\usepackage{graphicx}

\usepackage{epsfig}

\def\vb#1{\mbox{\boldmath$#1$}}
\def\pd#1#2{\frac{\partial #1}{\partial #2}}

\def\wh#1{\widehat{#1}}
\def\bdot{\,\vb{\cdot}\,}
\def\btimes{\,\vb{\times}\,}

\def\bhat{\wh{{\sf b}}}
\def\cal#1{\mathcal{#1}}

\def\exd{{\sf d}}

\newcommand{\bc}{\begin{center}}
\newcommand{\ec}{\end{center}}
\newcommand{\bt}{\begin{tabbing}}
\newcommand{\et}{\end{tabbing}} 
\newcommand{\be}{\begin{eqnarray*}}
\newcommand{\ee}{\end{eqnarray*}}
\newcommand{\bs}{\begin{slide}}
\newcommand{\es}{\end{slide}}

\begin{document}

\title{Comment on ``Geometric phase of the gyromotion for charged particles in a time-dependent magnetic field'' [Phys.~Plasmas {\bf 18}, 072505 (2011)]}

\author{Alain J.~Brizard$^{1}$ and Lo\"{i}c de Guillebon$^{2}$}
\affiliation{$^{1}$Department of Physics, Saint Michael's College, Colchester, VT 05439, USA \\ $^{2}$Centre de Physique Th\'{e}orique
Aix-Marseille Universit\'{e}, CNRS (UMR 7332), 13288 Marseille cedex 09 France} 

\begin{abstract}
The geometric analysis of the gyromotion for charged particles in a time-dependent magnetic field by J.~Liu and H.~Qin [Phys.~Plasmas {\bf 18}, 072505 (2011)] is reformulated in terms of the spatial angles that represent the instantaneous orientation of the magnetic field. This new formulation, which includes the equation of motion for the pitch angle, clarifies the decomposition of the gyroangle-averaged equation of motion for the gyrophase into its dynamic and geometric contributions.
\end{abstract}

\begin{flushright}
July 17, 2012
\end{flushright}

%\pacs{52.30.Gz, 52.65.Tt}

\maketitle

In a recent paper \cite{Liu_Qin}, Liu and Qin studied the dynamics of the gyrophase of a charged particle moving in a time-dependent (but uniform) magnetic field ${\bf B}(t) \equiv B_{x}(t)\,\wh{\sf x} + B_{y}(t)\,\wh{\sf y} + B_{z}(t)\,
\wh{\sf z}$. Liu and Qin then obtained an expression for the equation of motion for the gyrophase in terms of the Cartesian components $(B_{x}, B_{y}, 
B_{z})$ and their time derivatives. By using this Cartesian formulation, Liu and Qin demonstrated the existence of an anholonomic (i.e., path-dependent) geometric contribution to the gyrophase shift that depends on the evolution history of the magnetic field during a gyro-period.

In related previous work, Littlejohn \cite{RGL_81,RGL_88} showed that the Hamiltonian theory associated with guiding-center motion in an inhomogeneous magnetic field \cite{RGL_83} displays an anholonomic geometric phase associated with the gyrogauge vector ${\bf R} \equiv \nabla\wh{\sf e}_{1}\bdot\wh{\sf e}_{2}$ constructed from two basis unit vectors $\wh{\sf e}_{1}$ and $\wh{\sf e}_{2} \equiv \bhat\btimes\wh{\sf e}_{1}$ that span the plane that is locally perpendicular to the local magnetic unit vector $\bhat(t) \equiv {\bf B}(t)/B(t)$. Liu and Qin \cite{Liu_Qin}, on the other hand, showed that anholonomic geometric phases also exist in the problem of the gyromotion of charged particles (i.e., not guiding-centers) in a time-dependent magnetic field.

In the present Comment, we show how the analysis of Liu and Qin can be greatly simplified by using the polar angle $\theta$ and the azimuthal angle 
$\varphi$ describing the orientation of the spatially-uniform magnetic field \cite{RGL_88}, where $\cos\theta(t) \equiv B_{z}(t)/B(t)$ and $\tan\varphi(t) \equiv B_{y}(t)/B_{x}(t)$. We also clarify the role played by the pitch angle in the evolution of the gyrophase. We now reformulate the analysis of Liu and Qin \cite{Liu_Qin} concerning the case of time-dependent (but uniform) magnetic field ${\bf B}(t)$. 

First, we introduce the decomposition of the particle velocity ${\bf v}$ in terms of its pitch angle $\lambda$ (i.e., ${\bf v}\bdot\bhat = |{\bf v}|\,\cos\lambda$) and gyroangle $\zeta$ (i.e., $\partial{\bf v}/\partial\zeta = {\bf v}\btimes\bhat$):
\begin{equation}
{\bf v} \;\equiv\; v\;\left( \cos\lambda\;\bhat \;+\; \sin\lambda\;\wh{\sf c} \right),
\label{eq:v_def}
\end{equation}
where the unit vector $\wh{\sf c} \equiv \wh{\sf a}\btimes\bhat \equiv \partial\wh{\sf a}/\partial\zeta$ depends explicitly on 
$\zeta$ and the speed $v = |{\bf v}|$ is a constant of the motion. 

Next, we obtain the equations of motion for $\lambda$ and $\zeta$ by introducing the {\it magnetic} unit vectors \cite{RGL_88}
\begin{equation}
\left. \begin{array}{rcl}
\bhat & = & \cos\theta\;\wh{\sf z} \;+\; \sin\theta\;\wh{\rho} \\
\wh{\theta} & = & -\,\sin\theta\;\wh{\sf z} \;+\; \cos\theta\;\wh{\rho} \;=\; \partial\bhat/\partial\theta \\
\wh{\varphi} & = & -\,\sin\varphi\;\wh{\sf x} \;+\; \cos\varphi\;\wh{\sf y} \;=\; \partial\wh{\rho}/\partial\varphi
\end{array} \right\}.
\label{eq:b_theta_phi}
\end{equation}
Here, the ``radial'' vector $\bhat$ points in the direction of the magnetic field at each point in space (for a time-dependent uniform field, this direction is only a function of time). Since the unit vectors \eqref{eq:b_theta_phi} satisfy $\bhat \equiv \wh{\theta}\btimes\wh{\varphi}$, the unit vectors $\wh{\theta} \equiv \wh{\sf e}_{1}$ and $\wh{\varphi} \equiv \wh{\sf e}_{2}$ provide a possible choice of unit vectors $(\wh{\sf e}_{1}, \wh{\sf e}_{2})$ in the plane perpendicular to the unit magnetic vector $\bhat$ (i.e., tangent to the unit sphere). 

\section{Gyrogauge Geometry}

With the choice \eqref{eq:b_theta_phi} for the unit vectors $(\bhat,\wh{\sf e}_{1},\wh{\sf e}_{2})$, we construct the 
{\it magnetic} one-forms
\begin{equation}
\left. \begin{array}{rcl}
\exd\bhat\bdot\wh{\sf e}_{1} & = & \exd\theta \\
\exd\bhat\bdot\wh{\sf e}_{2} & = & \sin\theta\;\exd\varphi \\
\exd\wh{\sf e}_{1}\bdot\wh{\sf e}_{2} & = & \cos\theta\;\exd\varphi
\end{array} \right\},
\label{eq:db_123}
\end{equation}
where $\exd$ denotes an exterior derivative \cite{Flanders,Cartan}. Since the vectors $(\bhat,\wh{\sf e}_{1},\wh{\sf e}_{2})$ are orthogonal, we also have $\exd\wh{\sf e}_{i}\bdot\bhat = -\,\exd\bhat\bdot\wh{\sf e}_{i}$ and $\exd\wh{\sf e}_{j}\bdot\wh{\sf e}_{i} = -\,\exd\wh{\sf e}_{i}\bdot
\wh{\sf e}_{j}$, for $i,j = 1,2$. These definitions are not unique, however, since a rotation of the perpendicular unit vectors $(\wh{\theta},
\wh{\varphi}) \rightarrow (\wh{\theta}^{\prime},\wh{\varphi}^{\prime})$ about the $\bhat$-axis generated by the gyrogauge angle $\psi$:
\begin{equation}
\left( \begin{array}{c}
\wh{\theta}^{\prime} \\
\\
\wh{\varphi}^{\prime} \end{array} \right) \;\equiv\; \left( \begin{array}{cc}
\cos\psi & \sin\psi \\
 & \\
-\,\sin\psi & \cos\psi
\end{array} \right)\;\cdot\; \left( \begin{array}{c}
\wh{\theta} \\
\\
\wh{\varphi} \end{array} \right)
\label{eq:gyrogauge_rot}
\end{equation}
leads to the new one-forms
\begin{equation}
\left. \begin{array}{rcl}
\exd\bhat\bdot\wh{\sf e}_{1}^{\prime} & = & \cos\psi\;\exd\theta \;+\; \sin\theta\,\sin\psi\;\exd\varphi \;\equiv\; \omega_{2}\;
\exd t \\
\exd\bhat\bdot\wh{\sf e}_{2}^{\prime} & = & -\,\sin\psi\;\exd\theta \;+\; \sin\theta\,\cos\psi\;\exd\varphi \;\equiv\; 
-\,\omega_{1}\;\exd t \\
\exd\wh{\sf e}_{1}^{\prime}\bdot\wh{\sf e}_{2}^{\prime} & = & \cos\theta\;\exd\varphi \;+\; \exd\psi \;\equiv\; \omega_{3}\;
\exd t
\end{array} \right\}.
\label{eq:omega_123}
\end{equation}
Equation \eqref{eq:omega_123} shows the deep connection between the magnetic one-forms and the Eulerian angular frequencies $(\omega_{1},\omega_{2},
\omega_{3})$, which are defined in terms of the Euler angles $(\varphi,\theta,\psi - \pi/2)$. We can thus imagine an infinitesimaly thin symmetric top spinning about its axis of symmetry $\bhat \equiv \wh{\sf e}_{3}$ (at constant $\theta$ and $\varphi$ with $d\psi/dt \neq 0)$, which also undergoes precession (at constant $\theta$ with $d\varphi/dt \neq 0$) and nutation (at constant $\varphi$ with $d\theta/dt \neq 0$).

If we define the gyrogauge one-form 
\begin{equation}
{\cal R} \;\equiv\; \cos\theta\,\exd\varphi, 
\label{eq:R_def}
\end{equation}
and we denote its gyrogauge transformation as
\begin{equation}
{\cal R}^{\prime} \;\equiv\; {\cal R} \;+\; \exd\psi,
\label{eq:R_prime}
\end{equation}
then the property of gyrogauge invariance
\begin{equation}
\exd{\cal R}^{\prime} \;=\; \exd{\cal R}
\label{eq:gyrogauge_inv}
\end{equation} 
is guaranteed by the identity $\exd^{2} \equiv 0$ (corresponding to the vector identity $\nabla\btimes\nabla \equiv 0$). The gyrogauge one-form \eqref{eq:R_def} has a simple geometrical interpretation in terms of the solid-angle two-form 
\cite{Flanders}
\begin{equation}
\Phi \;\equiv\; \exd\theta\;\wedge\;\sin\theta\;\exd\varphi \;=\; -\;\exd\left(\cos\theta\frac{}{}\exd\varphi\right)
\;=\; -\;\exd{\cal R}.
\label{eq:area_2form}
\end{equation}
Hence, according to Eq.~\eqref{eq:gyrogauge_inv}, the solid-angle two-form \eqref{eq:area_2form} is a gyrogauge-invariant. In addition, using Stokes' Theorem \cite{Flanders}, the solid angle $\Omega \equiv \int_{\cal D}\,\Phi$ defined by the open surface ${\cal D}$ on the unit sphere is also expressed as
\begin{equation}
\Omega \;=\; -\;\int_{\cal D}\,\exd{\cal R} \;\equiv\; -\;\oint_{\partial{\cal D}}\;{\cal R},
\label{eq:Omega_Stokes}
\end{equation}
where $\partial{\cal D}$ denotes the boundary of ${\cal D}$. 

\section{Pitch-angle and gyroangle dynamics}

Next, we introduce the gyration unit vectors $\bhat \equiv \wh{\sf c}\btimes\wh{\sf a}$ in the plane perpendicular to $\bhat$:
\begin{eqnarray}
\wh{\sf a} & \equiv & \cos\zeta\;\wh{\theta} \;-\; \sin\zeta\;\wh{\varphi}, \label{eq:a_def} \\
\wh{\sf c} & \equiv & -\;\sin\zeta\;\wh{\theta} \;-\; \cos\zeta\;\wh{\varphi}. \label{eq:c_def}
\end{eqnarray}
These definitions are gyrogauge-invariant (i.e., $\wh{\sf a}^{\prime} = \wh{\sf a}$ and $\wh{\sf c}^{\prime} = \wh{\sf c}$) under the transformation $\zeta^{\prime} = \zeta + \psi$ and Eq.~\eqref{eq:gyrogauge_rot}. Using Eq.~\eqref{eq:v_def}, the equation of motion $d{\bf v}/dt = \omega_{\rm c}\,{\bf v}\btimes\bhat$ becomes
\begin{eqnarray}
0 & = & \omega_{\rm c}\,\sin\lambda\;\wh{\sf a} \;+\;\cos\lambda\;\left( \frac{d\bhat}{dt} \;+\; \frac{d\lambda}{dt}\;
\wh{\sf c} \right) \nonumber \\
 &  &+\; \sin\lambda\;\left( \frac{d\wh{\sf c}}{dt} \;-\; \frac{d\lambda}{dt}\;\bhat \right),
\label{eq:v_eq_abc}
\end{eqnarray}
where $\omega_{\rm c}(t) \equiv qB(t)/m$ denotes the time-dependent cyclotron frequency and, using 
Eqs.~\eqref{eq:a_def}-\eqref{eq:c_def}, we find
\begin{eqnarray}
\frac{d\bhat}{dt} & = & \frac{d\theta}{dt}\;\wh{\theta} \;+\; \sin\theta\;\frac{d\varphi}{dt}\;\wh{\varphi} \nonumber \\
 & = & \left( \cos\zeta\;\frac{d\theta}{dt} \;-\; \sin\zeta\,\sin\theta\;\frac{d\varphi}{dt} \right)\;
\wh{\sf a} \nonumber \\
 &  &-\; \left( \sin\zeta\;\frac{d\theta}{dt} \;+\; \cos\zeta\,\sin\theta\;\frac{d\varphi}{dt} \right)\;\wh{\sf c}, 
\label{eq:bhat_dot} \\
\frac{d\wh{\sf c}}{dt} & = & -\;\left( \frac{d\zeta}{dt} \;-\; \cos\theta\;\frac{d\varphi}{dt} \right)\;\wh{\sf a} \nonumber \\
 &  &+\; \left( \sin\zeta\;\frac{d\theta}{dt} \;+\; \cos\zeta\,\sin\theta\;\frac{d\varphi}{dt} \right)\;\bhat.
\label{eq:c_dot}
\end{eqnarray}
We note that these expressions can also be expressed as $d\bhat/dt = \omega_{1}\,\wh{\sf a} - \omega_{2}\,\wh{\sf c}$ and $d\wh{\sf c}/dt = \omega_{3}\,\wh{\sf a} + \omega_{2}\,\bhat$ in terms of the Eulerian angular frequencies $(\omega_{1},\omega_{2},\omega_{3})$ defined in Eq.~\eqref{eq:omega_123} with the substitution $\psi \rightarrow \pi/2 -\,\zeta$ (i.e., gyromotion occurs as a counter-rotation about $\bhat$). With these new definitions for the Eulerian frequencies, we easily recover the standard relations $\partial\omega_{1}/\partial\zeta = -\,\omega_{2}$, $\partial\omega_{2}/\partial\zeta = \omega_{1}$, and $\partial\omega_{3}/\partial\zeta = 0$.

Equation \eqref{eq:v_eq_abc} can be divided into two separate equations of motion:
\begin{eqnarray}
0 & = & \frac{d\lambda}{dt} \;+\; \frac{d\wh{\sf b}}{dt}\bdot\wh{\sf c}, 
\label{eq:lambda_bc} \\
0 & = & \omega_{\rm c} \;+\; \frac{d\wh{\sf c}}{dt}\bdot\wh{\sf a} \;+\; \cot\lambda\;\frac{d\wh{\sf b}}{dt}\bdot\wh{\sf a}. 
\label{eq:Omega_abc}
\end{eqnarray}
By using Eq.~\eqref{eq:bhat_dot}, Eq.~\eqref{eq:lambda_bc} yields the equation of motion for the pitch angle $\lambda$:
\begin{equation}
\frac{d\lambda}{dt} \;=\; \sin\zeta\;\frac{d\theta}{dt} \;+\; \cos\zeta\,\sin\theta\;\frac{d\varphi}{dt} \;\equiv\; \omega_{2}.
\label{eq:lambda_dot}
\end{equation}
While this equation is not considered by Liu and Qin \cite{Liu_Qin}, it plays an important role in the evolution of the gyrophase [see 
Eq.~\eqref{eq:zeta_dot_lambda} below]. We note that, since the angular velocities $d\theta/dt$ and $d\varphi/dt$ are gyroangle-independent (i.e., they represent the rate of change of the orientation of the magnetic field), the pitch-angle equation \eqref{eq:lambda_dot} satisfies $\langle d\lambda/dt\rangle = 0$ (which is valid for a uniform magnetic field), where $\langle\cdots\rangle$ denotes a gyroangle-average.

By using Eqs.~\eqref{eq:bhat_dot}-\eqref{eq:c_dot}, on the other hand, Eq.~\eqref{eq:Omega_abc} yields the equation of motion for the gyroangle $\zeta$:
\begin{eqnarray}
\frac{d\zeta}{dt} & = & \omega_{\rm c} \;+\; \cos\theta\;\frac{d\varphi}{dt} \nonumber \\
 &  &+\; \cot\lambda \left( \cos\zeta\;\frac{d\theta}{dt} \;-\; \sin\zeta\,\sin\theta\;\frac{d\varphi}{dt} \right) \nonumber \\
 & \equiv & \omega_{\rm c} \;+\; \cos\theta\;\frac{d\varphi}{dt} \;+\; \omega_{1}\;\cot\lambda.
\label{eq:zeta_dot}
\end{eqnarray}
Equation \eqref{eq:zeta_dot} corresponds exactly to Eq.~(7) of the paper \cite{Liu_Qin} by Liu and Qin, with the substitution $\zeta \rightarrow -\,\zeta - \pi/2$ and
\[ \frac{B_{z}}{B}\;\left(\frac{B_{x}\,\dot{B}_{y} - B_{y}\,\dot{B}_{x}}{B_{x}^{2} + B_{y}^{2}}\right) \;=\; \cos\theta\;
\frac{d\varphi}{dt}. \]
In Eq.~\eqref{eq:zeta_dot}, $\omega_{\rm c}$ is described as the {\it dynamical} term by Liu and Qin, while $\cos\theta\;d\varphi/dt$, which is clearly related to the gyrogauge one-form \eqref{eq:R_def}, is described as the {\it geometric} term. The last term on the right side of Eq.~\eqref{eq:zeta_dot} is described by Liu and Qin as an {\it adiabatic} term (because it is shown in the Appendix of Ref.~\cite{Liu_Qin} to be one order higher than the geometric term when the magnetic field evolves slowly compared to the gyration period). Instead, we use the pitch-angle equation \eqref{eq:lambda_dot} to write $\partial(d\lambda/dt)/\partial\zeta = \omega_{1}$ (once again valid for a uniform magnetic field), so that Eq.~\eqref{eq:zeta_dot} is written as
\begin{equation}
\frac{d\zeta}{dt} \;=\; \omega_{\rm c} \;+\; \cos\theta\;\frac{d\varphi}{dt} \;+\; \pd{}{\zeta}\left( \cot\lambda\;
\frac{d\lambda}{dt}\right).
\label{eq:zeta_dot_lambda}
\end{equation}
The third term in Eq.~\eqref{eq:zeta_dot} therefore appears as an exact gyroangle derivative, which disappears when 
Eq.~\eqref{eq:zeta_dot} is gyroangle-averaged: 
\begin{equation}
\left\langle\frac{d\zeta}{dt}\right\rangle \;=\; \omega_{\rm c} \;+\; \cos\theta\;\frac{d\varphi}{dt}.
\label{eq:zeta_dot_ave}
\end{equation}
We note that Eq.~\eqref{eq:zeta_dot_ave} is gyrogauge-invariant since, under a gyrogauge transformation generated by $\psi$, we have 
\[ \left( \left\langle\frac{d\zeta^{\prime}}{dt}\right\rangle,\; \frac{d\wh{\theta}^{\prime}}{dt}\bdot\wh{\varphi}^{\prime} \right) = \left( \left\langle\frac{d\zeta}{dt}\right\rangle + \frac{d\psi}{dt},\; \cos\theta\;\frac{d\varphi}{dt} + 
\frac{d\psi}{dt} \right), \]
which follows from the gyrogauge transformation \eqref{eq:R_prime}. Upon gyroangle-averaging, the equation of motion \eqref{eq:zeta_dot_ave} for the gyroangle is therefore decomposed in terms of dynamical and geometric terms only.

\section{Geometrical contributions to the gyrophase}

We now follow Ref.~\cite{Liu_Qin} and use Eq.~\eqref{eq:zeta_dot_ave} to calculate the averaged gyrophase shift, denoted 
$\langle\Delta\zeta\rangle$, in one gyration period $T \equiv 2\pi/\omega_{c}(t)$:
\begin{eqnarray}
\langle\Delta\zeta\rangle & \equiv & \int_{t}^{t + T}\left\langle\frac{d\zeta}{dt^{\prime}}\right\rangle\,dt^{\prime} = \langle\Delta\zeta_{\rm d}\rangle + \langle\Delta\zeta_{\rm g}\rangle[{\cal C}].
\label{eq:Delta_zeta}
\end{eqnarray}
The first term on the right side of Eq.~\eqref{eq:Delta_zeta} denotes dynamical gyrophase shift 
\begin{equation}
\langle\Delta\zeta_{\rm d}\rangle \;\equiv\; \int_{t}^{t + T}\,\omega_{\rm c}(t^{\prime})\;dt^{\prime}, 
\label{eq:Delta_zeta_dyn}
\end{equation}
which equals $2\pi$ for a time-independent magnetic field. The second term on the right side of Eq.~\eqref{eq:Delta_zeta}, on the other hand, denotes the geometrical gyrophase shift
\begin{eqnarray}
\langle\Delta\zeta_{\rm g}\rangle[{\cal C}] & \equiv & \int_{t}^{t + T}\cos\theta(t^{\prime})\;\frac{d\varphi}{dt^{\prime}}\;
dt^{\prime} \nonumber \\
 & = & \int_{\cal C}\;\cos\theta\;d\varphi \;=\; \int_{\cal C}\;{\cal R},
\label{eq:Delta_zeta_geo}
\end{eqnarray}
where the path ${\cal C}$ moves on the unit sphere from the initial point at $\theta(t)$ and $\varphi(t)$, to the final point at $\theta(t + T)$ and $\varphi(t + T)$. We note that the geometrical gyrophase shift \eqref{eq:Delta_zeta_geo} is path-dependent (i.e., it is anholonomic) since, by constructing the closed contour $\partial{\cal D} = {\cal C}_{1} - {\cal C}_{2}$ from two paths with identical end points, we find
\begin{eqnarray}
\langle\Delta\zeta_{\rm g}\rangle[{\cal C}_{1}] \;-\; \langle\Delta\zeta_{\rm g}\rangle[{\cal C}_{2}] & = & 
\oint_{\partial{\cal D}}\;{\cal R} \;=\; \int_{\cal D}\;\exd{\cal R}  \nonumber \\
 & = & -\;\Omega \;\neq\; 0,
\label{eq:zeta_path}
\end{eqnarray}
where $\Omega$ denotes the solid angle enclosed by the open surface ${\cal D}$ on the unit sphere.

Lastly, we introduce the time-scale ordering on the evolution of the magnetic field
\begin{equation}
\epsilon \;\equiv\; T\,\left|\frac{d\ln B}{dt}\right| \;\sim\; T\,\left|\frac{d\varphi}{dt}\right| \;\sim\; T\,\left|\frac{d\theta}{dt}\right| \;\ll\; 1,
\label{eq:epsilon_order}
\end{equation}
where the magnitude and direction of the magnetic field are assumed to change on the same slow time scale compared to the gyro-period $T$. By inserting this ordering in Eqs.~\eqref{eq:Delta_zeta_dyn}-\eqref{eq:Delta_zeta_geo}, we find the dynamical gyrophase shift
\begin{equation}
\langle\Delta\zeta_{\rm d}\rangle \;\simeq\; 2\pi \;+\; \pi\,\left(T\,\frac{d\ln B}{dt}\right)
\label{eq:zeta_dyn_epsilon} 
\end{equation}
and the geometrical gyrophase shift
\begin{eqnarray}
\langle\Delta\zeta_{\rm g}\rangle[{\cal C}] & \simeq & \frac{d\varphi}{dt}\;\int_{0}^{T}\;\cos[\theta(t + \tau)]\;d\tau
\nonumber \\
 & \simeq & \cos\theta\;\left(T\;\frac{d\varphi}{dt}\right),
\label{eq:zeta_geo_epsilon} 
\end{eqnarray} 
which is just $T$ times the instantaneous value of the geometric term in Eq.~\eqref{eq:zeta_dot_ave}, and is of order $\epsilon$. Finally, Liu and Qin compute the average value of the adiabatic term in Eq.~\eqref{eq:zeta_dot_lambda} and show that it is of order $\epsilon^2$. In our view, this contribution disappears upon gyroaveraging. 

\section{Summary} 

The formulation of the gyromotion in terms of the spatial angles relates the gyroangle dynamics with the motion of a spinning rigid body, through a natural appearance of the Euler frequencies. In addition, it emphasizes the role of the gyrogauge one-form $\mathcal R$ in the geometric interpretation of the gyromotion anholonomy. Lastly, the inclusion of the pitch-angle dynamics in our formulation shows that it is related to the adiabatic contribution to the gyroangle dynamics and it explains its adiabaticity. \\

We conclude this Comment with a few remarks concerning a general magnetic field that is space-time-dependent (details will be presented elsewhere). In this case, Eq.~\eqref{eq:zeta_dot_ave} is replaced with
\begin{equation}
\left\langle\frac{d\zeta}{dt}\right\rangle \;=\; \omega_{\rm c} \;+\; \sigma \;+\; \left( \tau\;\frac{ds}{dt} +\frac{1}{2}\,\frac{d\chi}{dt}\right),
\label{eq:zeta_dot_ave_gen}
\end{equation}
where $\exd\wh{\sf e}_{1}\bdot\wh{\sf e}_{2} \equiv \sigma\,\exd t + {\bf R}\bdot\exd{\bf x}$ \cite{RGL_81}. Here, two additional contributions appear. The first one involves the (Frenet-Serret) torsion of the magnetic-field line $\tau \equiv \bhat\bdot{\bf R} = (\partial\wh{\sf e}_{1}/\partial s)
\bdot\wh{\sf e}_{2}$, with $v_{\|} \equiv ds/dt$ used in Eq.~\eqref{eq:zeta_dot_ave_gen}. The second one involves the twist of the magnetic-field lines 
$\tau_{\rm m} = \bhat\bdot\nabla\btimes\bhat \equiv d\chi/ds$, defined as the rate of rotation (denoted by the angle $\chi$) of a nearby field line about the magnetic field line represented by $\bhat$ \cite{Moffatt_Ricca,Berger_Prior}, with $v_{\|}\,\tau_{\rm m} = d\chi/dt$ used in 
Eq.~\eqref{eq:zeta_dot_ave_gen}. We note that the torsion and the magnetic twist can be comparable in some magnetic geometries \cite{footnote}. The torsion contributes to an anholonomic (path-dependent) geometric gyrophase shift $\langle\Delta\zeta_{\tau}\rangle[{\cal C}] \equiv \int_{\cal C}\tau\,ds$ while the magnetic twist contributes an holonomic (path-independent) geometric gyrophase shift $\langle\Delta\zeta_{\chi}\rangle[{\cal C}] \equiv 
\frac{1}{2}\,\int_{\cal C}d\chi = \frac{1}{2}\,\Delta\chi$.


\begin{thebibliography}{99}
\bibitem{footnote} For example, in a screw-pinch cylindrical geometry $(r,\theta,z)$ described by $\bhat_{0} = \cos\Theta\,
\wh{\sf z} + \sin\Theta\,\wh{\theta}$, $\bhat_{1} = -\,\wh{\sf r}$, and $\bhat_{2} = -\,\cos\Theta\,\wh{\theta} + \sin\Theta\,
\wh{\sf z}$, where $\Theta \equiv \Theta(r)$, we find the magnetic curvature $\kappa = r^{-1}\,\sin^{2}\Theta$, the Frenet-Serret torsion $\tau = r^{-1}\,\cos\Theta\,\sin\Theta$, and the magnetic torsion (twist) $\tau_{\rm m} = \Theta^{\prime} + \tau$.
\end{thebibliography}
\end{document}